\documentclass[12pt,a4paper]{article}
\usepackage{amsmath} 
\usepackage{mathtools}
\usepackage{amsfonts}
\usepackage{amssymb}
\usepackage{graphicx} 
\usepackage {anysize} 
\usepackage{natbib}
\title{\textbf{A physically-based analytical model to describe effective excess charge for streaming potential generation in water saturated porous media}}

\author{L. Guarracino$^{1,2,+,*}$ and D. Jougnot$^{3,+}$}

\begin{document}
\marginsize{2.5cm}{2cm}{2cm}{2cm}

\maketitle

\noindent
(1) Consejo Nacional de Investigaciones Cient\'ificas y T\'ecnicas, Facultad de Ciencias Astron\'omicas y Geof\'isicas, Universidad Nacional de La Plata, La Plata, Argentina                    

\noindent
(2) Facultad de Ciencias Naturales y Museo, Universidad Nacional de La Plata, La Plata, Argentina 

\noindent
(3) Sorbonne Universit\'e,  CNRS, EPHE, UMR 7619 Metis, F-75005, Paris, France 

\noindent
(+){Both authors contributed equally to this publication}

\noindent
(*)Corresponding author: luisg@fcaglp.unlp.edu.ar

\vskip 8cm

\noindent
This paper has been published in \textit{Journal of Geophysical Research -- Solid Earth}, please cite as: \\

\noindent
\textit{\textbf{L. Guarracino, D. Jougnot (2018) A physically-based analytical model to describe effective excess charge for streaming potential generation in water saturated porous media, Journal of Geophysical Research - Solid Earth, 123(1), 52-65, doi:10.1002/2017JB014873.}}

\newpage

\paragraph{Keypoints}

\begin{itemize}
\item Derivation of a physically-based analytical model to determine the effective excess charge
\item Mechanistic explanation for the empirical dependence between effective excess charge and permeability
\item The new model reproduces experimental data for different media and a broad range of ionic concentrations
\end{itemize}

\vskip 2cm

\begin{abstract}
Among the different contributions generating self-potential, the streaming potential is of particular interest in hydrogeology for its sensitivity to water flow. Estimating water flux in porous media using streaming potential data relies on our capacity to understand, model, and upscale the electrokinetic coupling at the mineral-solution interface. Different approaches have been proposed to predict streaming potential generation in porous media. One of these approaches is the flux averaging which is based on determining the excess charge which is effectively dragged in the medium by water flow. 
In this study, we develop a physically-based analytical model to predict the effective excess charge in saturated porous media using a flux-averaging approach in a bundle of capillary tubes with a fractal pore-size distribution.
The proposed model allows the determination of the effective excess charge as a function of pore water ionic concentration and hydrogeological parameters like porosity, permeability and tortuosity. The new model has been successfully tested against different set of experimental data from the literature. One of the main findings of this study is the mechanistic explanation to the empirical dependence between the effective excess charge and the permeability that has been found by several researchers. The proposed model also highlights the link to other lithological properties and it is able to reproduce the evolution of effective excess charge with electrolyte concentrations.
\end{abstract}

\newpage

%
%

\section{Introduction}

The self-potential (SP) method is a passive geophysical method based on the measurements of the electrical field which is naturally generated in the subsurface. The SP method was first proposed by Robert Fox in 1830 \citep{fox1830electromag} and it is considered as one of the oldest geophysical methods. Although SP data are relatively easy to measure, the extraction of useful information is a non-trivial task since the recorded signals are a superposition of different SP components. In natural porous media, signals are mainly generated by electrokinetic (water flux) and electrochemical (ionic fluxes or redox reactions) phenomena. For more details of this method and for an overview of all possible SP sources we refer to \citet{revil2013self}.

\begin{figure}[h]
\centering
\includegraphics[scale=0.65]{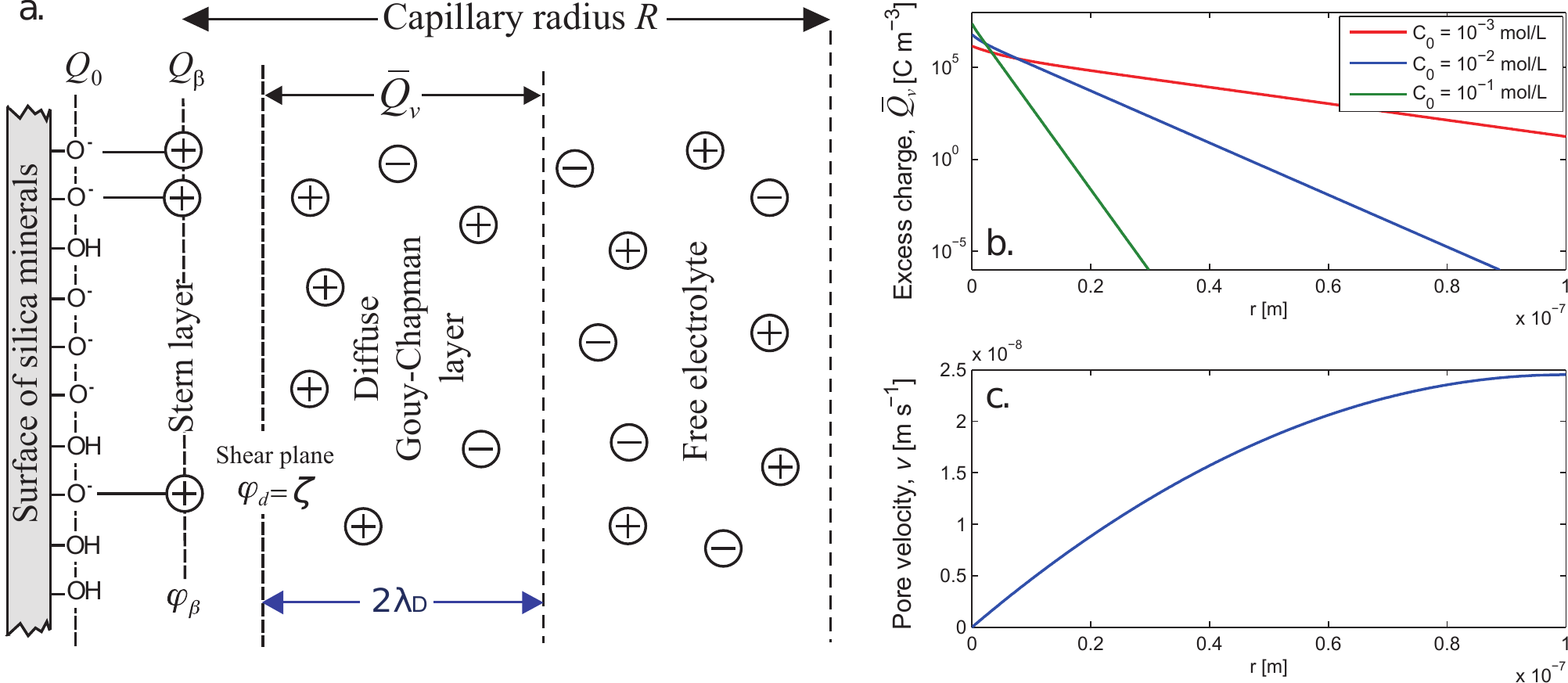}
\caption{(a) Scheme of the electrical double layer for a given capillary of radius $R$. Distribution of the excess charge (b) and the pore vater velocity (c) from the shear plane ($r = 0$ m) to the center of the capillary ($r = R = 10^{-7}$ m). The excess charge is calculated using (\ref{qv}) for different ionic concentrations of NaCl and the \cite{jaafar2009measurement} model for the $\zeta$-potential. Figure modified from \cite{jougnot2015monitoring}.}
\label{Fig:TLM}
\end{figure}

In the present study, we focus on the electrokinetic (EK) contribution to the SP, that is the part of the signal generated from the water flow in porous media (often referred to as streaming potential). The surface of the minerals that constitute porous media is generally electrically charged, creating an electrical double layer (EDL) containing an excess of charge that counterbalances the charge deficiency of the mineral surface \citep[see][]{Hunter1981,leroy2004triple}. Figure \ref{Fig:TLM}a shows how the EDL is composed: a Stern layer that contains only counterions coating the mineral with a very limited thickness, and a diffuse layer that contains both counterions and coions but with a net excess charge. The shear plane, that can be approximated as the limit between the Stern layer and diffuse layer \citep[e.g.][]{leroy2004triple}, separates the mobile and immobile part of the water molecules when subjected to a pressure gradient. This plane is characterized by an electrical potential called $\zeta$-potential \citep[see][]{Hunter1981}. When the water flows through the pore it drags a fraction of the excess charge that give rise to a streaming current and an electrical potential field.

The EK phenomenon has been studied experimentally and theoretically for more than a century due its relevance in many practical applications in various fields (e.g., microfluidic, chemical engineering, fluid mechanics). In hydrogeophysics, its interest relies on the fact that SP data is sensitive to groundwater flow. Indeed, groundwater flow monitoring can otherwise be performed locally using rather intrusive methods (i.e. hydraulic head measurements between boreholes or in situ sensors, heat or chemical tracer evolution from a single well). However, given that SP only provides indirect measurements (electrical potential) of water flux, it is necessary to rely on a theoretical framework and petrophysical relationships that can predict the EK phenomenon at macroscopic scales. 

Two main approaches to simulate streaming current generation in fully saturated porous media can be found in the literature. The classical approach relies on the use of the coupling coefficient, which is a rock-dependent property that relates the difference of hydraulic pressure to the difference in electrical potential. This property has first been described experimentally \citep[e.g.][]{quincke1859ueber, dorn1880ueber} and later quantified by \cite{helmholtz1879studien} and \cite{smoluchowski1903contribution}. The so-called Helmholtz-Smoluchowski coupling coefficient has been developed from a capillary-tube model and its final expression does not depend on the geometrical properties of the porous medium. Therefore, it has been used for any kind of medium under the assumption that the electrical conductivity of the mineral surface could be neglected. When it is not the case, alternative formula have been proposed by several researchers \cite[e.g.,][]{revil1999streaming,glover2010streaming}. Various model using capillaries to predict the coupling coefficient under saturated and partially saturated conditions can be found in the literature \cite[e.g.][]{rice1965electrokinetic, ishido1981experimental, jackson2010multiphase, jackson2012validity, thanh2017fractal}.
The second approach to simulate streaming current generation is more recent and focuses on the excess charge that is dragged by the water flow. The first reference to this approach in the English literature can be found in \citet{kormiltsev1998three} and more detailed theoretical frameworks can be found in  \citet{revil2004constitutive}, \citet{revil2007electrokinetic} and \citet{revil2016transport1, revil2016transport2}. In this approach, the coupling parameter is the excess charge that is effectively dragged by water in the porous media. 
Then, the streaming current and streaming potential distribution can be generated by multiplying the effective excess charge by the water velocity. Note that both approaches describe the same physics; the main difference between them relies on the parameter (coupling coefficient or excess charge density) used to describe the electrokinetic coupling  between fluid flow and streaming potential generation.

In this work, we focus on the excess charge approach in the framework proposed by \cite{Sill1983SP} and modified by \cite{kormiltsev1998three} and \cite{revil2007electrokinetic}, where SP signals can be directly related to the water velocity:
\begin{equation}
\nabla \cdot \left( \sigma \nabla \varphi \right) = \nabla \cdot  \left( \hat{Q}_v \textbf{v}_D \right)
\label{Eq_Poisson}
\end{equation}
where, $\sigma$ is the electrical conductivity (S m$^{-1}$), $\varphi$ the electrical potential (V) and $\hat{Q}_v$ the excess charge density (C m$^{-3}$) effectively dragged by the water flow at a given Darcy velocity $\textbf{v}_D$ (m s$^{-1}$). From Eq. (\ref{Eq_Poisson}), it becomes clear that determining the effective excess charge is crucial to predict streaming potential signal in hydrosystems and to infer water flux from SP measurements.

Experimental evidences has shown that the effective excess charge depends on the porous medium permeability \citep{titov2002electrokinetic, jardani2007tomography, boleve2012dyke}.
The first relationship to estimate the effective excess charge from permeability can be found in \cite{titov2002electrokinetic} (Fig. 2 of their paper). Some years later, \citet{jardani2007tomography} proposed a more precise empirical relationship that has been proven to be very useful as it decreases the number of variables to be estimated and provides good estimates of efective excess charge for different types of porous media. This relationship has been used in many studies involving electrokinetic phenomena, such as dam leakage \cite[e.g.][]{boleve2009preferential, ikard2014characterization}, surface-groundwater interaction \cite[e.g.][]{linde2011self}, seismoelectric studies \cite[e.g.][]{mahardika2012waveform, jougnot2013seismoelectric,revil2015seismoelectric,monachesi2015analytical}, hydraulically-active fracture identification \citep[e.g.][]{roubinet2016streaming} and permeability field characterization \cite[e.g.][]{jardani2009stochastic, soueid2014hydraulic}. However, Eq. (\ref{jardani}) has been obtained regardless of pore water composition and other hydraulic parameters of porous media.
Unfortunately, this relationship does not take into account the dependence of the EDL on the ionic concentration of the electrolyte \cite[see the discussion in][]{jougnot2015monitoring}. 

The most rigorous approach to construct an accurate description of any phenomena at macroscopic scales is to start with a pore-scale description and then apply an upscaling method to the microscopic equations \citep{Bernabe2015}. In this study, the porous media is conceptualized as a bundle of capillary tubes with a fractal pore-size distribution. The porosity and permeability of the porous media are estimated using an equivalent medium theory. 
This geometrical description of the porous media and upscaling procedure have been successfully used to describe water flow in fractured media \citep{guarracino2006, guarracino2009}, the evolution of multiphase flow properties during mineral dissolution \citep{guarracino2014fractal} and relations between hydraulic parameters \citep{guarracino2007estimation}. On the other hand, the effective excess charge in a single tube can be calculated from the radial distributions of excess charge and water velocity. Then the effective excess charge at the macroscopic scale is estimated using the flux-averaging technique proposed by \citet{jougnot2012derivation}. In that paper, they show that by combining these hydraulic and EK properties, a closed-form expression for the effective excess charge can be obtained in terms of permeability, porosity, tortuosity, ionic concentration, $\zeta$-potential, and Debye length. The dependence of the developed model on ionic concentration, permeability, porosity and  also grain size is tested using different set of experimental data. It is also shown that the proposed model allows to derive from physical concepts the empirical relation between effective excess charge and permeability found by \citet{jardani2007tomography}.

\section{Theoretical development}

The proposed model is based on the macroscopic description of effective excess charge in porous media from the upscaling of pore-size flow and electrokinetic phenomena. The porous medium is conceptualized as an equivalent bundle of water-saturated capillary tubes with a fractal law distribution of pore sizes. First, we derive expressions for the most important macroscopic hydraulic parameters (porosity and permeability). Then, we obtain an approximate expression for the excess charge in a single tube and, from this result, we estimate the effective excess in the porous media in terms of the hydraulic parameters.    
 
To derive both hydraulic and electrokinetic properties we consider as representative elementary volume (REV) a cylinder of radius $R_{REV}$ (m) and lenght $L$ (m). The pores are assumed to be circular capillary tubes with radii varing from a minimum pore radius $R_{min}$ (m) to a maximum pore radius $R_{max}$ (m).

The cumulative size-distribution of pores whose radii are greater than or equal to $R$ (m) is assumed to obey the following fractal law  \citep{tyler1990fractal,yu2003permeabilities,guarracino2014fractal}:
\begin{equation}
N(R) = \left(\frac{R_{REV}}{R}\right)^{D},
\label{nr}
\end{equation}
where $D$ is the fractal dimension of pore and $0<R_{min}\leq R\leq R_{max}< R_{REV}$. By using the Sierpinski carpet (a classical fractal object) \citet{tyler1990fractal} show that the fractal dimension of (\ref{nr}) ranges from 1 to 2, with the highest values associated with the finest textured soils. Capillary tube models with this type of fractal pore-size distributions have been successfully used to describe macroscopic process and hydraulic properties in different porous media \cite[e.g.][]{yu2003permeabilities, guarracino2014fractal,soldi2017}.

Note that the cumulative number of pores given by (\ref{nr}) decreases with the increase of pore radius $R$, then differentiating (\ref{nr}) with respect to $-R$ we can obtain the number of pores whose radii are in the infinitesimal range $R$ to $R+dR$:
\begin{equation}
dN = D R_{REV}^{D} R^{-D-1} dR.
\label{dn} 
\end{equation}

\subsection{Hydraulic properties}

The porosity $\phi$ of the REV defined above can be straightforward computed from its definition:
\begin{equation}
\phi = \frac{\text{Volume of pores}}{\text{Volume of REV}}= \frac{ \int_{R_{min}}^{R_{max}} V_p(R) dN }{\pi R_{REV}^2 L},
\label{poro1}
\end{equation}
where $V_p(R)=\pi R^2 l$ (m$^3$) is the volume of a single tube of radius $R$ and length $l$ (m). 

Substituting (\ref{dn}) into (\ref{poro1}) and solving the definite integral we obtain
\begin{equation}
\phi = \frac{\tau D}{ R_{REV}^{2-D} (2-D)} \left( R_{max}^{2-D} - R_{min}^{2-D} \right),
\label{poro2}
\end{equation}
where $\tau=l/L$ is the dimensionless hydraulic tortuosity of the capillary tubes \citep{scheidegger1958physics}. Note that the model assumes a single value of $\tau$ for all capillary tubes, this value must be considered as a mean tortuosity value of all tube sizes.    

In order to obtain both the permeability and the excess charge of the REV we need to describe the water flow in a single tube. For a laminar flow rate, the velocity distribution inside the tube can be described by the Poiseuille model \citep{bear88}:
\begin{equation}
v(R,r) = \frac{\rho_w g}{4 \eta \tau} \left[ R^2 - (R-r)^2 \right] \frac{\Delta h}{L},
\label{q1}
\end{equation}
where $r$ (m) is the distance from the pore wall ($r=0$) to the center of the tube ($r=R$), $\rho_w$ the water density (kg/m$^3$), $g$ the gravitational acceleration (m/s$^2$), $\eta$ the dynamic viscosity (Pa s), and $\Delta h$ the pressure head drop across the REV (m). The average velocity $\overline{v}$ (m/s) in the capillary tube has the following expression:
\begin{equation}
\overline{v}(R) = \frac{\rho_w g}{8 \eta \tau}  R^2  \frac{\Delta h}{L}.
\label{q2}
\end{equation}

The total volumetric flow through the REV $V_Q$ (m$^3$/s) is the sum of the volumetric flow rates of all individual tubes. According to (\ref{q2}) and (\ref{dn}) $V_Q$ can be computed as follows:
\begin{equation}
V_Q =  \int_{R_{min}}^{R_{max}} \overline{v}(R) \pi R^2 dN = \frac{\rho_w g}{8 \eta \tau} \frac{\pi D R_{REV}^D}{4-D} \left(R_{max}^{4-D}-R_{min}^{4-D}\right) \frac{\Delta h}{L}.
\label{Q}
\end{equation} 
On the basis of Darcy's law (macroscopic scale), $V_Q$ can also be expressed as
\begin{equation}
V_Q  = \pi R_{REV}^2 \frac{\rho_w g}{\eta} k \frac{\Delta h}{L},
\label{QD}
\end{equation} 
$k$ being the intrinsic permeability of the porous media (m$^2$). 

Then, combining  (\ref{Q}) and (\ref{QD}) we obtain the following expression for permeability $k$ in terms of the geometrical parameters of the porous media: 
\begin{equation}
k  =  \frac{D}{ 8 \tau (4-D) R_{REV}^{2-D} } \left(R_{max}^{4-D}-R_{min}^{4-D}\right).
\label{perm}
\end{equation}

It is important to remark that a similar equations for $\phi$ and $k$ have been recently derived by \citet{soldi2017} assuming constrictive capillary tubes. In the limit case of straight tubes both expressions for $\phi$ and $k$ are identical (see equations (11) and (15) of their paper).

For most porous media it can be assumed that $R_{min} << R_{max}$ \citep{yu2001}. Then, equations (\ref{poro2}) and (\ref{perm}) can be reduced to
\begin{equation}
\phi = \frac{\tau D}{(2-D) R_{REV}^{2-D} } R_{max}^{2-D},
\label{poro3}
\end{equation}
\begin{equation}
k  =  \frac{D}{ 8 \tau (4-D) R_{REV}^{2-D} } R_{max}^{4-D}.
\label{perm2}
\end{equation}

Finally, combining (\ref{poro3}) and (\ref{perm2}) we obtain a simple relationship to estimate permeabilty from porosity
\begin{equation}
k  = \gamma  \phi^{ \frac{4-D}{2-D}},
\label{permporo}
\end{equation}
where $ \gamma=\frac{R_{REV}^2}{8 \tau (4-D)} \left(\frac{2-D}{\tau D} \right)^{\frac{4-D}{2-D}}$. Note that for $D=1$ the exponent of the porosity is 3 and (\ref{permporo}) is equivalent to Kozeny's equation \citep{kozeny1927kapillare}.    

\subsection{Electrokinetic properties}

Since electrokinetic phenomenon is caused by the coupling of fluid flow and charge distribution at pore scale, the magnitude of this phenomenom will be mainly determined by the macroscopic hydraulic and electrical properties of the porous medium. Based on the previous description of hydraulic properties we will compute the effective excess charge density $\hat{Q}_v^{REV}$ carried by the water flow in the REV (C/m$^3$). 
The effective excess charge density, also called dynamic excess charge depending on the authors and symbolized by $\hat{Q}_v$ or $\bar{Q}_v^{eff}$, has to be distinguished from the other excess charge densities contained in the pore space: $Q_v$ the total excess charge density (C/m$^3$), which includes all the charges from the Stern and diffuse layers of the EDL, and $\bar{Q}_v$ the excess charge located only in the diffuse layer (C/m$^3$) (Fig. \ref{Fig:TLM}a) \cite[see the discussion in][]{revil2017dependence}.
It is important to remark that in this study we do not consider the charges that are located in the Stern layer as they are fixed on the pore wall and do not contribute to the streaming current.

We start this derivation by defining the excess charge distribution $\bar{Q}_v$ in a capillary tube saturated by a binary symetric 1:1 electrolyte (e.g., NaCl) as follows
\begin{equation}
\bar{Q}_v (r)  = N_A e_0 C^0 \left[  e^{-\frac{e_0\psi(r)}{k_B T}} -  e^{\frac{e_0\psi(r)}{k_B T}}\right],
\label{qv}
\end{equation}
where $N_A$ is the Avogadro's number (mol$^{-1}$), $e_0$ the elementary charge (C), $C^0$ the ionic concentration far from the mineral surface (mol/m$^3$), $\psi$ the local electrical potential in the pore water (V), $k_B$ the Boltzman constant (J/K), and $T$ is the absolute temperature (K). For the thin double layer assumption (i.e., the thickness of the double layer is small compared to the pore size) the local electrical potential can be expressed  \citep{Hunter1981}:
\begin{equation}
\psi (r)  = \zeta e^{-\frac{r}{l_D}},
\label{z}
\end{equation}
\begin{equation}
l_D =  \sqrt{ \frac{\epsilon k_B T}{2 N_A C^0 e_0^2} },
\label{ld}
\end{equation}
where $\zeta$ (V) is the  $\zeta$-potential on the shear plane (Fig. \ref{Fig:TLM}a), $l_D$ the Debye lenght (m), and $\epsilon$ the water dielectric permittivity (F/m).

As proposed by \citet{jougnot2012derivation}, the effective excess charge density $\hat{Q}^{R}_v$ carried by the water flow in a single tube of radius $R$ is defined by
\begin{equation}
\hat{Q}^{R}_v  = \frac{1}{\overline{v}(R) \pi R^2} \int_{A} \overline{Q}_v(r) v(R,r) dA,
\label{qv1}
\end{equation} 
being $A$ the cross sectional area of the tube (m$^2$). Using a polar coordinate system with the pole located in the center of the tube, equation (\ref{qv1}) becomes
\begin{equation}
\hat{Q}^{R}_v = \frac{2}{\overline{v}(R) R^2} \int_0^R \overline{Q}_v(r) v(R,r) (R-r) dr.
\label{qv2}
\end{equation}  

In order to obtain a closed-form analytical expression of $\hat{Q}^{R}_v$ we approximate the exponential terms of (\ref{qv}) by a four-term Taylor serie:
\begin{equation}
e^{\pm \frac{e_0\psi(r)}{k_B T}}  = 1  \pm \frac{e_0\psi(r)}{k_B T} + \frac{1}{2} \left( \frac{e_0\psi(r)}{k_B T} \right)^2 \pm \frac{1}{6} \left( \frac{e_0\psi(r)}{k_B T} \right)^3.
\label{taylor}
\end{equation}

Substituting (\ref{taylor}) in (\ref{qv}) and solving (\ref{qv2}) we obtain
\begin{equation}
\begin{array}{ll}
\hat{Q}^{R}_v = &-\frac{8 N_A e_0^2 C^0 \zeta}{k_B T (R/l_D)^4} \left\{ 6 -e^{-\frac{R}{l_D}} \left[ \left(\frac{R}{l_D}\right)^3 + 3\left(\frac{R}{l_D}\right)^2 + 6\left(\frac{R}{l_D}\right) +6 \right]  \right\} \\
& +\frac{24 N_A e_0^2 C^0 \zeta}{k_B T (R/l_D)^3} \left\{ 2 -e^{-\frac{R}{l_D}} \left[ \left(\frac{R}{l_D}\right)^2 + 2\left(\frac{R}{l_D}\right) + 2 \right]  \right\} \\
& -\frac{16 N_A e_0^2 C^0 \zeta}{k_B T (R/l_D)^2} \left\{ 1 -e^{-\frac{R}{l_D}} \left[ \left(\frac{R}{l_D}\right) + 1 \right]  \right\} \\
& -\frac{4 N_A e_0^4 C^0 \zeta^3}{3 (k_B T)^3 (3R/l_D)^4}  \left\{ 6 -e^{-\frac{3R}{l_D}} \left[ \left(\frac{3R}{l_D}\right)^3 + 3\left(\frac{3R}{l_D}\right)^2 + 6\left(\frac{3R}{l_D}\right) +6 \right]  \right\} \\
& +\frac{4 N_A e_0^4 C^0 \zeta^3}{(k_B T)^3 (3R/l_D)^3} \left\{ 2 -e^{-\frac{3R}{l_D}} \left[ \left(\frac{3R}{l_D}\right)^2 + 2\left(\frac{3R}{l_D}\right) + 2 \right]  \right\} \\
& -\frac{8 N_A e_0^4 C^0 \zeta^3}{3 (k_B T)^3 (3R/l_D)^2} \left\{ 1 -e^{-\frac{3R}{l_D}} \left[ \left(\frac{3R}{l_D}\right) + 1 \right]  \right\}.
\end{array}
\label{qvana}
\end{equation}

For the thin double layer assumption (i.e., the thickness of the double layer is small compared to the pore size) we consider $l_D << R$ and (\ref{qvana}) can be reduced to
\begin{equation}
\hat{Q}^{R}_v = \frac{8 N_A e_0 C^0}{(R/l_D)^2}  \left[ - 2 \frac{e_0 \zeta}{k_B T}  -\left(\frac{e_0 \zeta}{3 k_B T} \right)^3 \right].
\label{qvap}
\end{equation}

\begin{figure}[h]
\centering
\includegraphics[]{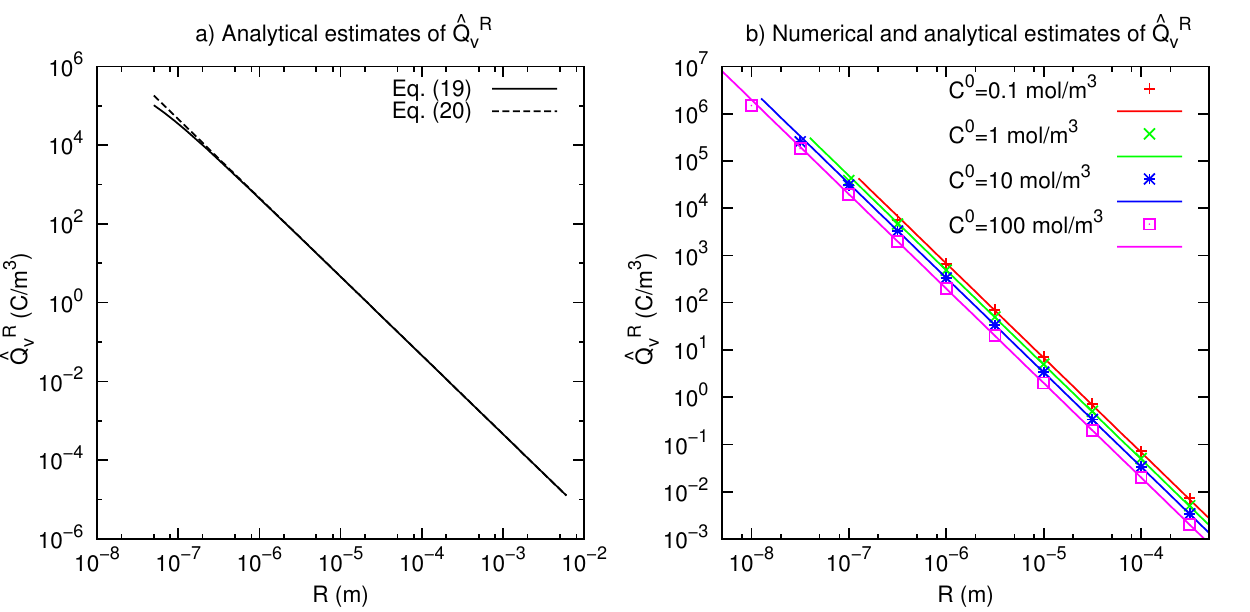}
\caption{ (a) Analytical estimates of $\hat{Q}^{R}_v$ using equations (\ref{qvana}) and (\ref{qvap}) for a ionic concentration $C^0=1$ mol/m$^3$; (b) numerical (points from Eq. \ref{qv1}) and analytical (lines from Eq. \ref{qvap}) estimates of $\hat{Q}^{R}_v$ for different concentration values ($R>5 l_D$). The $\zeta$-potential values are computed using (\ref{zeta}).}
\label{fig2}
\end{figure}

Figure \ref{fig2} (a) shows the effective excess charge $\hat{Q}^{R}_v$ predicted by (\ref{qvana}) and (\ref{qvap}) for a ionic concentration $C^0=1$ mol/m$^3$. Note that even though the number of terms of equation (\ref{qvap}) is drastically reduced, both equations predict similar values of $\hat{Q}^{R}_v$.    
In order to test the general validity of (\ref{qvap})under the thin double layer assumption, we compare approximate values of $\hat{Q}^{R}_v$ with exact values obtained by the numerical solution of (\ref{qv2}) assuming pore-sizes $R$ greater that 5 Debye lenghts. Figure \ref{fig2} (b) presents the goodness of the fit for different values of ionic concentration $C^0$. From the analysis of Figure \ref{fig2}, we conclude that equation (\ref{qvap}) predicts fairly well the effective excess charge in capillary tubes for a wide range of radius and ionic concentration values.  

The effective excess charge $\hat{Q}^{REV}_v$ carried by the water flow in the REV is defined by
\begin{equation}
\hat{Q}^{REV}_v  = \frac{1}{v_D \pi R_{REV}^2} \int_{R_{min}}^{R_{max}} \hat{Q}^{R}_v(R) \overline{v}(R) \pi R^2 dN,
\label{qrev1}
\end{equation} 
where $v_D=\frac{\rho_w g}{\eta} k \frac{\Delta h}{L}$ is the Darcy's velocity (m/s) (macroscopic scale). Substituting (\ref{qvap}), (\ref{q2}) and (\ref{dn}) in (\ref{qrev1}) and assuming $R_{min} << R_{max}$ yields
\begin{equation}
\hat{Q}^{REV}_v  =  8 N_A e_0 C^0 \left[ - 2 \frac{e_0 \zeta}{k_B T}  -\left(\frac{e_0 \zeta}{3 k_B T} \right)^3 \right]  \frac{4-D}{2-D} \left(\frac{l_D}{R_{max}} \right)^2.
\label{qrev2}
\end{equation}
Finally, combining (\ref{poro3}), (\ref{perm2}) and (\ref{qrev2}) we obtain the following expression for $\hat{Q}^{REV}_v$:
\begin{equation}
\hat{Q}^{REV}_v  =  N_A e_0 C^0 l_D^2  \left[ - 2 \frac{e_0 \zeta}{k_B T}  -\left(\frac{e_0 \zeta}{3 k_B T} \right)^3 \right]  \frac{ 1}{\tau^2} \frac{\phi}{k}.
\label{qrev3}
\end{equation}
The above equation constitutes the main result of this paper. 
Note that (\ref{qrev3}) predicts the effective excess charge density in terms of both macroscopic hydraulic parameters (porosity, tortuosity and permeability) and electrokinetic parameters (ionic concentration, $\zeta$-potential and Debye lenght). This equation gives insight into the role of macroscopic hydraulic parameters and it can be considered a starting point for designing non-invasive methods to monitoring groundwater flow using self-potential measurements.     

\begin{figure}[h]
\centering
\includegraphics[]{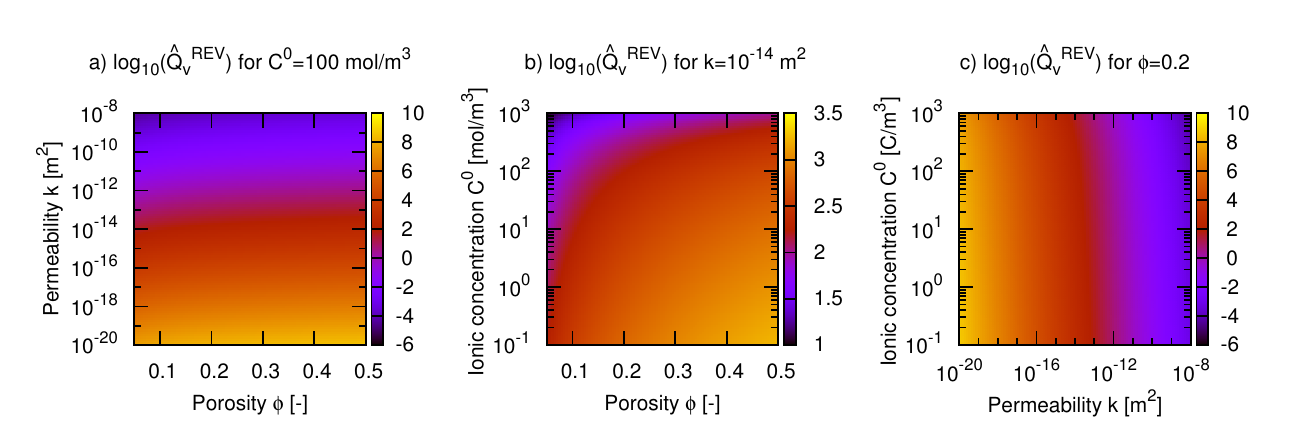}
\caption{ Parametric analysis of effective excess charge $\hat{Q}^{REV}_v$: (a) sensitivity to porosity and permeability for a fixed value of ionic concentration, (b) sensitivity to porosity and ionic concentration for a fixed value of permeability, (c) sensitivity to permeability and ionic concentration for a fixed value of porosity.}
\label{fig3}
\end{figure}

In order to study the role of porosity $\phi$, permeability $k$, and ionic concentration $C^0$ on effective excess charge, we perform a parametric analysis of equation (\ref{qrev3}). 
The ionic concentration dependence of $\zeta$-potential is assumed to obey the relation proposed by \citet{pride1991electrokinetic}:
\begin{equation}
\zeta(C^0)  = a + b \text{log}_{10}(C^0),
\label{zeta}
\end{equation}
where $a$ and $b$ are fitting parameters. For this study we use the parameter values obtained by \citet{jaafar2009measurement} on silicate-based materials for NaCl brine $a$=-6.43 mV and $b$=20.85 mV. 

Figure \ref{fig3} summarizes the parametric analysis of $\log_{10}\left(\hat{Q}^{REV}_v\right)$ for the following ranges of variability of ionic concentration $C^0$, permeability $k$ and porosity $\phi$: $10^{-1}$ mol/m$^3$ $\le C^0 \le$ $10^{3}$ mol/m$^3$, $10^{-20}$ m$^2$ $\le k \le$ $10^{-8}$ m$^2$ and $0.05$ $\le \phi \le$ $0.5$.
Figure \ref{fig3}a shows the effect of porosity and permeability on $\hat{Q}^{REV}_v$ for a fixed value of ionic concentration ($C^0=100$ mol/m$^3$). It can be observed that $\hat{Q}^{REV}_v$ is strongly determined by permeability while porosity only produces a slightly increase of $\hat{Q}^{REV}_v$ values. Figure \ref{fig3}b shows the effect of porosity and ionic concentration on $\hat{Q}^{REV}_v$. As shown in this panel, these parameters can change $\hat{Q}^{REV}_v$ values in two orders of magnitude for a fixed value of permeability ($k=10^{-14}$ m$^2$). An increase of $\hat{Q}^{REV}_v$ can be observed when ionic concentration decreases and porosity increases. Finally, Figure \ref{fig3}c shows the role of permeability and ionic concentration on $\hat{Q}^{REV}_v$ for a fixed value of porosity ($\phi=0.2$). It can be observed that the effect of permeability on $\hat{Q}^{REV}_v$ is much more significant than ionic concentration.   

From this parametric analysis we can conclude that effective excess charge $\hat{Q}^{REV}_v$ is highly sensitive to permeability values. However, porosity and ionic concentration can modify $\hat{Q}^{REV}_v$ values in two orders of magnitude for a given value of permeability \citep[see discussions in][]{jougnot2012derivation,jougnot2015monitoring}. The increase of porosity or the decrease of ionic concentration produce an increase of $\hat{Q}^{REV}_v$ values.

\section{Comparison with the empirical relationship proposed by \citet{jardani2007tomography}}

\begin{figure}[h]
\centering
\includegraphics[scale=0.8]{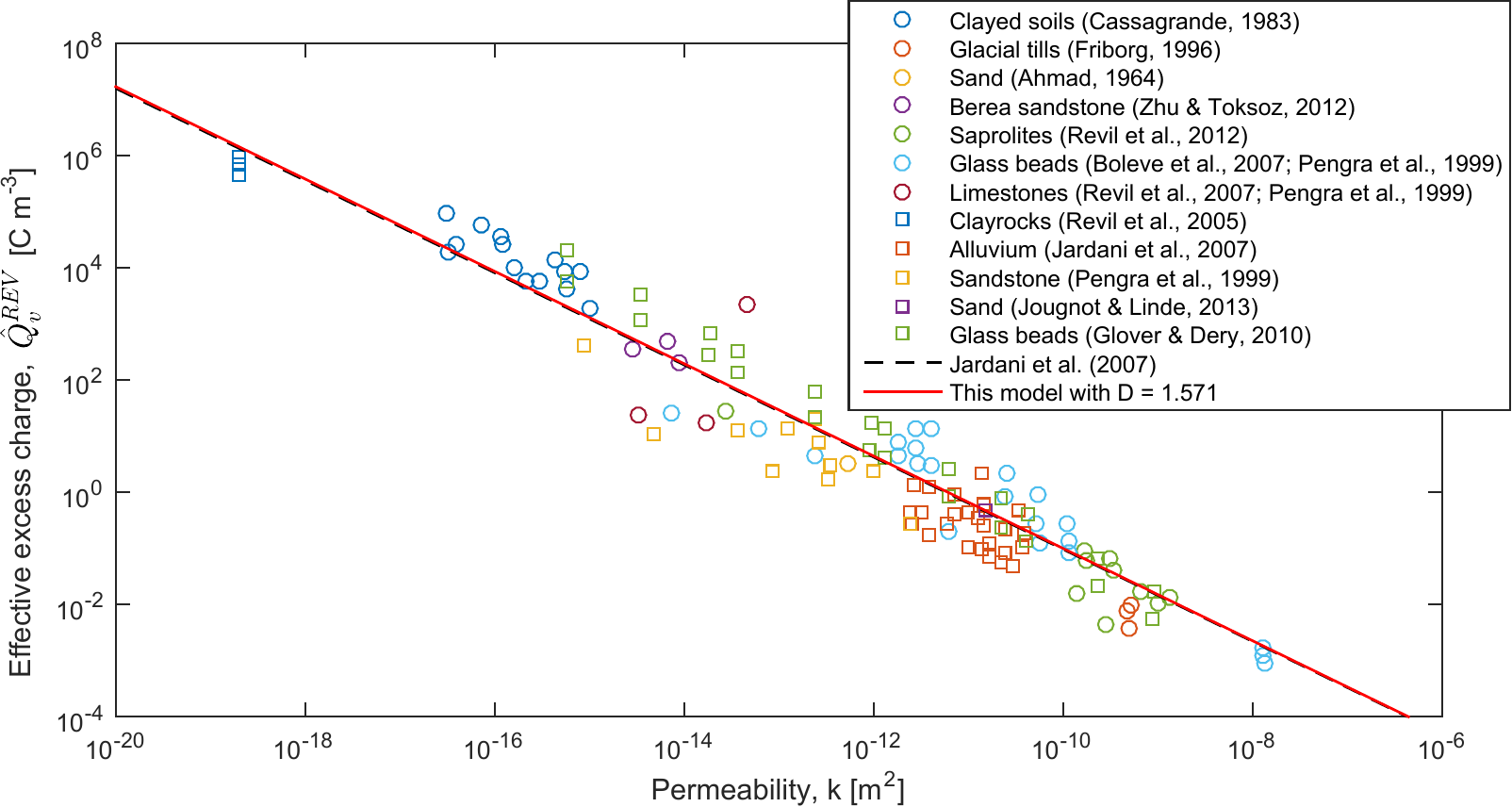}
\caption{Comparison between the effective excess charge $\hat{Q}^{REV}_v$ as a function of the permeability $k$. Symbols represent experimental data from literature for different lithologies. Solid lines are the fit of the proposed model (with $D=1.571$) and the \citet{jardani2007tomography} empirical relationship (\ref{jardani}). }
\label{Qv_k}
\end{figure}

The empirical relationship to estimate the effective excess charge $\hat{Q}^{REV}_v$ from permeability $k$ proposed by \citet{jardani2007tomography} reads as follows:
\begin{equation}
\text{log}_{10}(\hat{Q}^{REV}_v)  = A_1 + A_2\text{log}_{10}(k),
\label{jardani}
\end{equation}
where $ A_1=-9.2349$ and  $ A_2= -0.8219$ are constant values obtained by fitting (\ref{jardani}) to a large set of experimental data that includes various lithologies and ionic concentrations (Fig. \ref{Qv_k}). Note that this equation has been obtained regardless of pore water composition and other hydraulic parameters of porous media. 

The empirical relationship (\ref{jardani}) can be derived from the proposed equation (\ref{qrev3}). Using (\ref{permporo}) we obtain the following expression for porosity in term of permeability: $\phi=(k/\gamma)^{ \frac{2-D}{4-D}}$. Replacing this expression in (\ref{qrev3}) and taking the logarithm on both sides of the resulting equation, we obtain exactly (\ref{jardani}) but with the following constants:
\begin{equation}
\begin{array}{ll}
A_1  =  \text{log}_{10} \left\{ \frac{N_A e_0 C^0}{\gamma^{\frac{2-D}{4-D}}}  \left[ - 2 \frac{e_0 \zeta}{k_B T}  -\left(\frac{e_0 \zeta}{3 k_B T} \right)^3 \right]  \left(\frac{l_D}{\tau} \right)^2   \right\}, \\
A_2  = - \frac{2}{4-D}.
\end{array}
\label{jardani2}
\end{equation}
According to our model, the $\text{log}_{10}(\hat{Q}^{REV}_v)$-intercept $A_1$ mainly depends on chemical and interface parameters while the slope $A_2$ only depends on the fractal dimension of the pore size distribution ($1<D<2$). Note that the predicted range of the slope is $-1<A_2<-0.666$ and that the value obtained by \citet{jardani2007tomography} $ A_2= -0.8219$ corresponds to the fractal dimension $D=1.567$.   

Figure \ref{Qv_k} shows the fit of (\ref{jardani2}) to an extensive set of $\hat{Q}^{REV}_v$ data determined by several authors \citep{ahmad1964laboratory, cassagrande1983stabilization, friborg1996experimental, pengra1999determination,  revil2005characterization, revil2007electrokinetic, revil2012petrophysical,  boleve2007streaming, jardani2007tomography, glover2010streaming, zhu2012formation, jougnot2013electrodes}. The best fit of our model is obtained for fractal dimension $D=1.571$. Note that this single value of $D$ can fit a wide range of soil textures, then it can be considered a mean value of all pore size distributions.

It is worth mentioning that $\hat{Q}^{REV}_v$ data displayed on Fig. \ref{Qv_k} are obtained from electrokinetic coupling coefficient $C^{EK}$ values (V/Pa) using the following equation (\cite{revil2004constitutive, revil2007electrokinetic}):
\begin{equation}
\hat{Q}^{REV}_v = - \dfrac{C^{EK} \sigma \eta}{k},
\label{def_C_EK}
\end{equation}
where $\sigma$ (S/m) is the electrical conductivity, $\eta$ (Pa s) the dynamic viscosity, and $k$ (m$^2$) the permeability. The values of $\sigma$ and $k$ of each sample have been measured independently.

If $D = 1.527$ allows us to retrieve the exact same trend as \citet{jardani2007tomography}, we are left with a cloud of values that spread around the proposed model. To test our model with more accuracy, we focus on a selection of experimental data where all the parameters have been measured in the following section.

\section{Application to laboratory data}

\subsection{Effect of salinity}

In order to analyze the effect of salinity on $\hat{Q}^{REV}_v$ we test the proposed model (\ref{qrev3}) with laboratory data obtained by \citet{pengra1999determination}. These authors performed an exhaustive petrophysical characterization of a collection of rock and glass bead samples and they measured the electrokinetic coupling coefficient $C^{EK}$ for different NaCl brine concentrations. For this test $\hat{Q}^{REV}_v$ values are obtained from $C^{EK}$ using (\ref{def_C_EK}). The only parameter of(\ref{qrev3}) which is not measured by \citet{pengra1999determination} is the hydraulic tortuosity $\tau$. Thus, we fit this parameter using a least square algorithm. The ionic concentration dependence of $\zeta$-potential is assumed to obey (\ref{zeta}). 

\begin{figure}[h]
\centering
\includegraphics[]{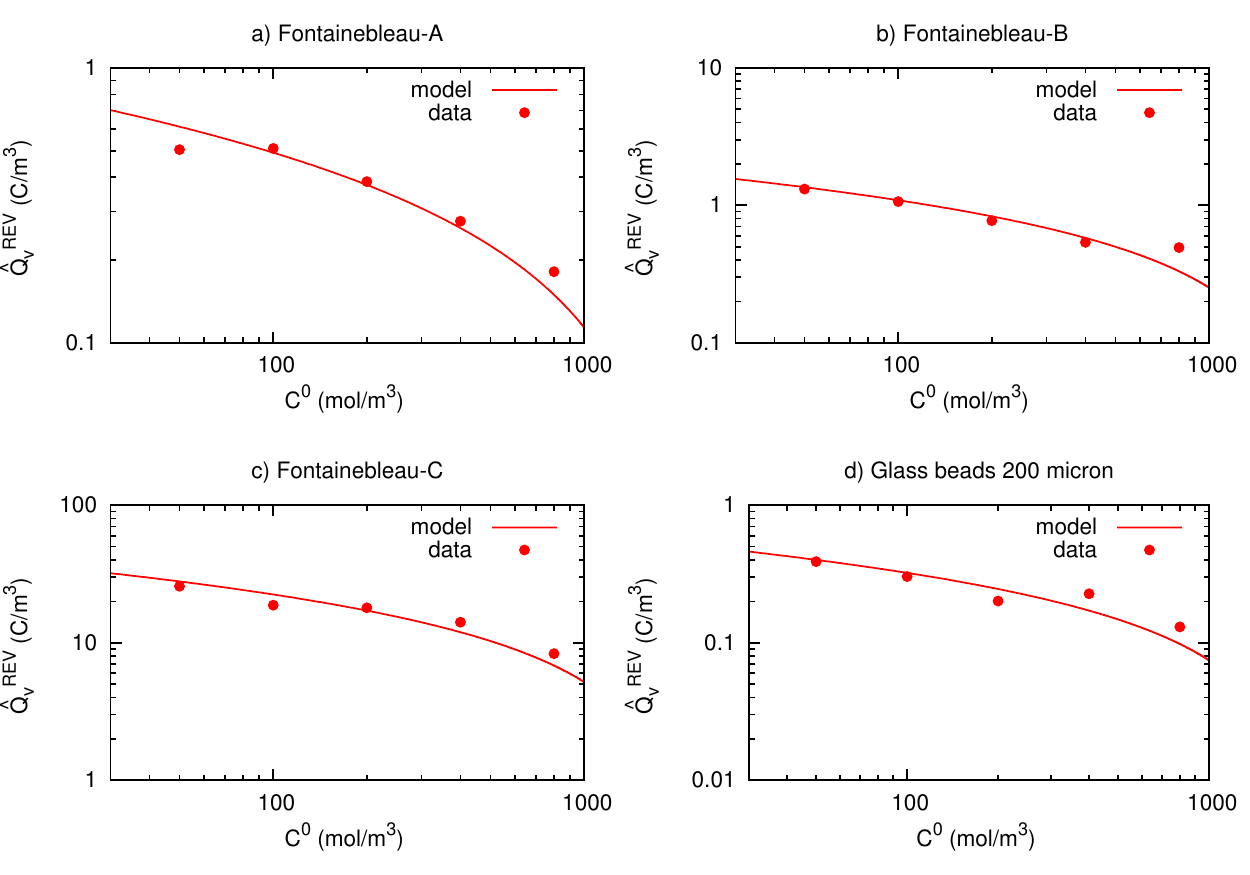}
\caption{Comparison between measured values of $\hat{Q}^{REV}_v$ and predicted values using the proposed model (Eq. \ref{qrev3}) for different ionic concentrations. Experimental data are obtained from \citet{pengra1999determination}. Hydraulic parameters of soil samples: a) $\phi=0.223$, $k=2.36 \times 10^{-12} \text{m}^2$, $\tau=1.95$, b) $\phi=0.168$, $k=9.09 \times 10^{-13} \text{m}^2$, $\tau=1.83$, c) $\phi=0.067$, $k=5.63 \times 10^{-15} \text{m}^2$, $\tau=3.24$, d) $\phi=0.298$, $k=5.07 \times 10^{-12} \text{m}^2$, $\tau=1.90$.  Tortuosity values are obtained by fitting (\ref{qrev3}) to experimental data.}
\label{fig5}
\end{figure}

Figure \ref{fig5} shows the fit of (\ref{qrev3}) to experimental values of effective excess charge $\hat{Q}^{REV}_v$ measured at different ionic concentrations $C^0$ for 3 sandstones samples of different permeabilities and 1 fused glass bead sample. The fitted values of tortuosity and measured values of porosity and permeability of each sample are listed in the figure caption. Experimental data show that $\hat{Q}^{REV}_v$ decreases with the increase of ionic concentration, and this behaviour can be adequately described by the proposed model. The decrease of $\hat{Q}^{REV}_v$ with the increase of $C^0$  was also predicted and discussed by \citet{jougnot2015monitoring}.

\subsection{Effect of porosity}

\begin{figure}[h]
\centering
\includegraphics[]{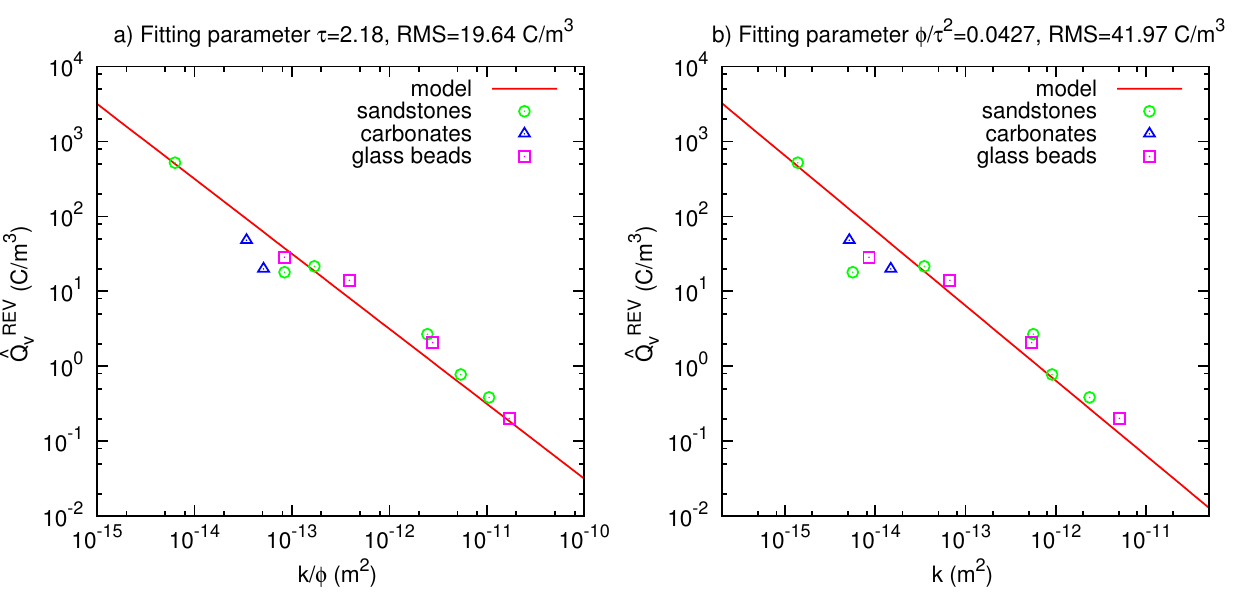}
\caption{Comparison between measured values of $\hat{Q}^{REV}_v$ vs $k/\phi$ and predicted values using (\ref{qrev3}) for $C_0= 0.2$ mol/L.}
\label{fig6}
\end{figure}

To test the effect of porosity on $\hat{Q}^{REV}_v$ estimates we fit the proposed model to experimental values of $\hat{Q}^{REV}_v$ in terms of both $k$ and $k/\phi$ obtained by \citet{pengra1999determination} for $C^0=0.2$ mol/L in different porous media. Figure \ref{fig6}a shows the fit of (\ref{qrev3}) to experimental data in terms of $k$. The fitting parameter is the ratio $\phi/\tau^2$ and the RMS of the fit is 41.97 C/m$^3$. Figure \ref{fig6}b shows the fit of the model to the same experimental values of $\hat{Q}^{REV}_v$ but in terms of $k/\phi$. In this case the fitting parameter is $\tau^2$ and the RMS is reduced to 19.64 C/m$^3$. Even though more experimental evaluations of the model are needed, this test shows that the inclusion of porosity improves the estimate of $\hat{Q}^{REV}_v$.

\subsection{Effect of grain size}

Among the available data in the SP literature, \citet{glover2010streaming} studied the effect of grain size on the electrokinetic coupling coefficient $C^{EK}$. These authors measured $C^{EK}$ on 12 packs of glass bead having different grain diameters $d$ ($d \in [1 , 990] \times 10^{-6}$ m) at two pore water salinities: $C^0 = 2 \times 10^{-4}$ and $2 \times 10^{-3}$ mol/L.

To test the performance of our model for different grain sizes, we used the measurements of \citet{glover2010streaming} to predict the effective excess charge. For each grain size, we used the permeability and mercury porosity listed in their Table 1. For ionic concentrations $C^0 = 2 \times 10^{-4}$ and $2 \times 10^{-3}$ mol/L we use, respectively, the mean values $\zeta =$ -71.62 and -24.89 mV of $\zeta$-potentials calculated for all the samples (given in their Table 2). Note that the glass bead diameter should not influence the interface properties of the quartz mineral, this is an artifact due to surface conductivity effects \cite[see also ][]{leroy2012double, li2016influence}. 

In order to compare our model prediction with the measurements, we compute the coupling coefficient $C^{EK}$ using (\ref{def_C_EK}) and the electrical conductivity model proposed by \citet{glover2010streaming}:
\begin{equation}
\sigma = \dfrac{1}{F} \left( \sigma_w + \dfrac{4 m F \Sigma_S}{d} \right) ,
\label{sig_glover}
\end{equation}
where $\sigma_w$ is the electrical conductivity of pore water (calculated from $C^0$, see \citep{sen1992influence}); $\Sigma_S$, the glass beads surface conductance; $m$, the cementation index; and $F = \phi^{-m}$, the formation factor of the bead samples. The only parameter which is not provided in the study is the hydraulic tortuosity $\tau$.

\begin{figure}[h]
\centering
\includegraphics[scale=0.5]{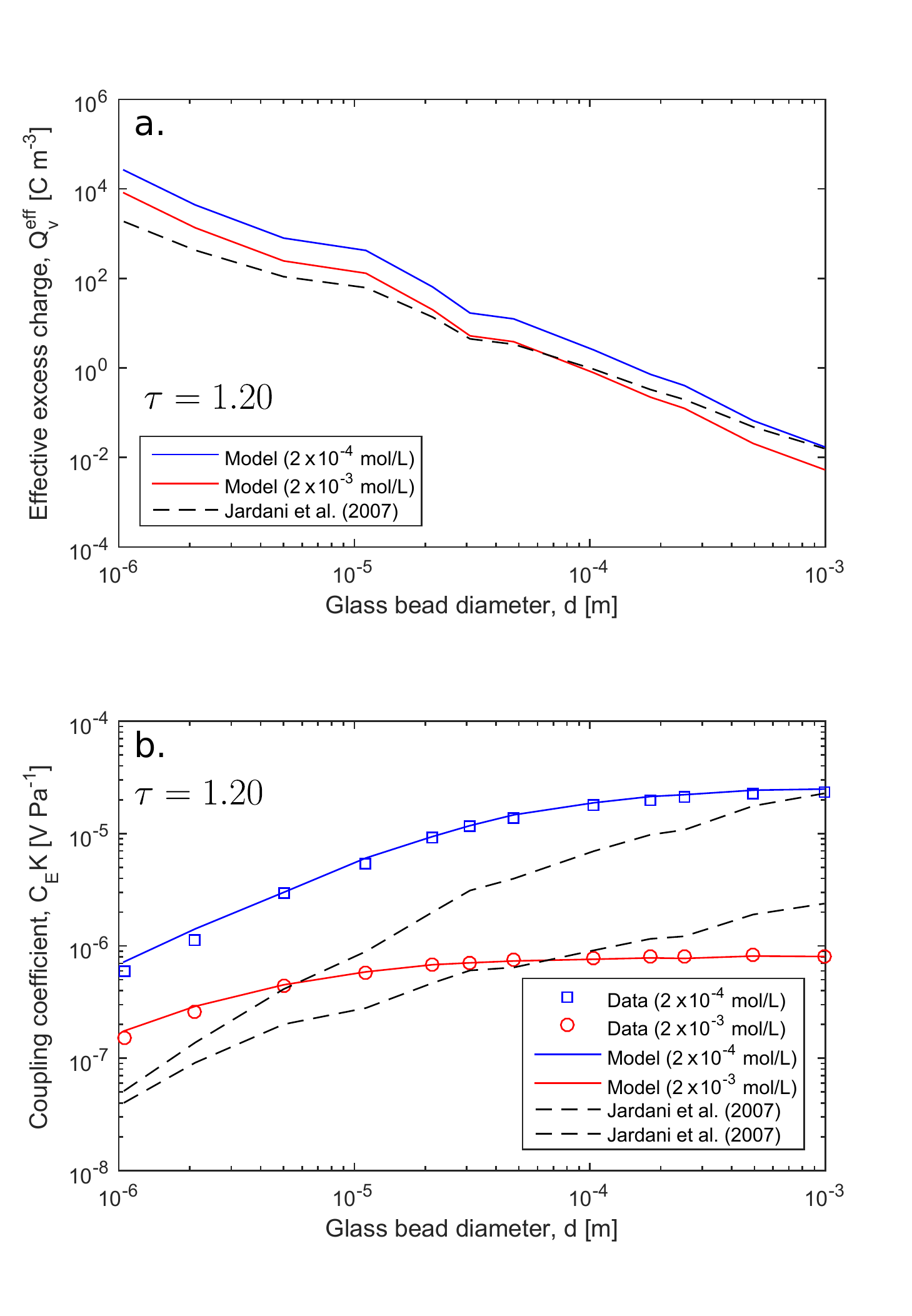}
\caption{(a) Model prediction of $\hat{Q}^{REV}_v$ for glass beads pack with different diameters at two different ionic concentrations: $C^0 = 2 \times 10^{-4}$ and $2 \times 10^{-3}$ mol/L in blue and red, respectively. (b) Comparison between the measured and predicted coupling coefficients when using the proposed model with $\tau =1.20$. Note that the black dashed lines are the \cite{jardani2007tomography} empirical relationship prediction (Eq. \ref{jardani}) for the effective excess charge in (a) and the coupling coefficient in (b).}
\label{GrainSize}
\end{figure}

Figure \ref{GrainSize}a shows the predicted effective excess charge for the two ionic concentrations as a function of the glass bead diameter and $\tau=1.2$ (this tortuosity has been chosen as it gives the best fit for all the data presented on Fig. \ref{GrainSize}b at once). We also plot in the figure the prediction of \cite{jardani2007tomography} empirical relationship which only depends on permeability $k$. 
Figure \ref{GrainSize}b illustrates the pretty good fit between the coupling coefficient measurements and predicted values of our model for both pore water salinities. Note that $\tau = 1.20$ is the only fitting parameter of the model and this parameter allows to describe all measured data (different glass bead sizes and two ionic concentration values). It is worth mentioning that we could improve the fit between the proposed model and experimental data by adjusting $\tau$ for each glass bead diameter (especially for $d \ll 5 \times 10^{-6}$ m), but we prefer keeping the number of fitting parameters as small as possible.

\section{Discussion and conclusion}

The present study is focused on the estimate of the effective excess charge $\hat{Q}^{REV}_v$ in fully saturated porous media, a key parameter to understand and model the streaming potential generation. Based on physical and geometrical concepts we derive a closed-form analytical expression to estimate $\hat{Q}^{REV}_v$ from electrokinetic and hydraulic parameters. The mathematical development involves up-scaling procedures at pore and REV scales, similar to the numerical up-scaling proposed by \citet{jougnot2012derivation} for partially saturated soils. The extension of the present model to unsaturated conditions is not straightforward as a relationship between permeability and saturation needs to be derived for the fractal pore size distribution.

The first step of this study consists in the up-scaling of EK properties for a binary symetric 1:1 electrolyte from the EDL scale to the capillary tube or pore scale (i.e., from few nm to possibly several mm). The approximate expression (\ref{qvap}) relates effective excess charge $\hat{Q}^{R}_v$ to the radius of the tube $R$ and chemical parameters (ionic concentration $C^0$, $\zeta$-potential and Debye length $l_D$). The accuracy of this equation has been tested to be correct as long as the electrical double layer from the walls of the capillary do not overlap ($R > 5 l_D$), i.e. under the thin double layer assumption (Fig. \ref{fig2}). An extension of this model for EDL overlapping could be obtained by truncation of the diffuse layer \cite[e.g.][]{goncalves2007introducing}.

The second step consists in the up-scaling of both hydraulic and EK properties from the pore scale (capillary tube) to the REV scale. 
The REV of porous media is conceptualized as a bundle of capillary tubes with a fractal pore-size distribution. Fractal models have been proven very useful to obtain different hydraulic properties of rocks and soils \citep{tyler1990fractal, yu2003permeabilities,guarracino2007estimation, guarracino2014fractal,soldi2017}. In this study, the fractal distribution approach allows us to obtain simple expressions for porosity $\phi$ (\ref{poro3}) permeability $k$ (\ref{perm2}) and effective excess charge $\hat{Q}_v^{REV}$ (\ref{qrev2}) in terms of the fractal dimension $D$, the maximum pore radius $R_{max}$ and the REV radius $R_{REV}$. Finally, combining these macroscopic properties we obtain the closed-form expression (\ref{qrev3}) to estimate $\hat{Q}_v^{REV}$ from hydraulic ($\phi$, $k$, $\tau$) and pore water chemistry ($C^0$, $l_D$, $\zeta$) parameters.  

The proposed model is derived from physical and geometrical concepts and provides a mechanistic framework for understanding the role of hydraulic parameters on $\hat{Q}_v^{REV}$. In particular, the model corroborates the empirical relationship (\ref{jardani}) of \citet{jardani2007tomography} which relies on an increasing number of data sets with different lithologies and it is used by many researchers. 
It is important to remark that the link between effective excess charge and permeability is the subject of debates in the scientific community \cite[e.g.][]{jouniaux2016review, zyserman2017dependence, revil2017dependence} and our model provides a theoretical justification to this link.
However, in this study we show that it would be better to link $\hat{Q}_v^{REV}$ to the ratio of permeability and porosity ($k/\phi$) instead of $k$ (see Fig. \ref{fig6}).

Another limitation of the relationship proposed by \citet{jardani2007tomography} is that it does not explicitly depend on the pore-water chemistry, i.e. on the $\zeta$-potential, Debye length and ionic concentration. Indeed, the newly proposed model takes into account the ionic concentration and its $\zeta$-potential dependence and is therefore able to reproduce laboratory data from \cite{pengra1999determination} where all the parameters but one (tortuosity) have been estimated independently.

The proposed model shows that $\hat{Q}^{REV}_v$ contains information on the lithology as predicted by \cite{revil2013self} (p. 64). Indeed, three petrophysical parameters are explicitly identified in the model: the permeability, $k$, the porosity, $\phi$, and the hydraulic tortuosity, $\tau$. It is worth to emphazise here that the relation between hydraulic and electrical tortuosity is not straightforward \cite[e.g.][]{clennell1997tortuosity}. In the present work, we test the simple model of \cite{winsauer1952resistivity} to determine the hydraulic tortuosity used in Eq. (\ref{qrev3}) based on parameters that can be measured electrically \cite[see ][ for a discussion]{clennell1997tortuosity}:
\begin{equation}
(\tau_e)^2 = F \phi,
\label{tau_e}
\end{equation}
where $\tau_e$ is the hydraulic tortuosity determined electrically from the formation factor and the porosity. Table \ref{table1} show the comparison between the fitted tortuosity $\tau$ to the one predicted by (\ref{tau_e}). One can see that predicted and fitted tortuosities fall fairly close one to an other, which is a promising preliminary result. However, further work is needed to establish a better relation between these electrical and hydraulic properties, therefore to link the effective excess charge $\hat{Q}^{REV}_v$ to electrical conductivity.

\begin{table}
\caption{Comparison between fitted and predicted hydraulic tortuosities}
\begin{tabular}{l c c c}
\hline 
Material & $\tau$ from fit & $\tau_e$ from Eq. (\ref{tau_e}) \\ 
\hline 
\textit{Glass beads packs from }\cite{glover2010streaming} \\
Mean value over 12 packs  & 1.20  & 1.24 \\ 
\hline 
\textit{Samples from }\cite{pengra1999determination}\\
Fontainebleau A & 1.95 & 1.59 \\ 
Fontainebleau B & 1.83 & 1.82 \\ 
Fontainebleau C & 3.24 & 3.03 \\ 
Glass beads (200 $\mu$m) & 1.90 & 1.64 \\
\hline
\label{table1}
\end{tabular}
\end{table}

The proposed model represents a major step forward in understanding the links between hydraulic and electrokinetic parameters in the framework of the excess charge approach. The model provides a theoretical basis for the empirical relationship with the medium permeability and describes the dependence of the excess charge on other petrophysical parameters. The simplicity of the model and its excellent fit to different set of experimental data opens-up new possibilities for a broader use of the SP method in hydrogeophysics studies based on the increasingly popular excess charge approach. The analytical development of a model for partially saturated porous media using the presented approach will be the natural next step to pursue this work.

%
%
%
%
%
%
%

\subsection*{Acknowledgments}

The data used to test the proposed model is listed in the references.
This research is partially supported by Universidad Nacional de La Plata and Consejo Nacional de Investigaciones Cient\' ificas y T\'ecnicas (Argentina). The authors strongly thank the editor and the three reviewer for their nice and really constructive comments.

\end{document}